\documentclass[doublecol]{epl2} 
\usepackage{epsf,graphicx,amssymb,subfig}
\usepackage[dvipsnames]{xcolor}
\usepackage{float}
\usepackage{amssymb}
\usepackage{graphicx}
\definecolor{violet}{rgb}{0.5,0,0.5}

\title{Intruder mobility in a vibrated granular packing}
\author{Rim Harich, Thierry Darnige, Evelyne Kolb, Eric Cl\'ement}
\shortauthor{Harich \etal}
\institute{Laboratoire de Physique et M\'ecanique des Milieux
  H\'et\'erog\`enes (UMR 7636 et Universit\'es Paris 6 et Paris 7), 10
  rue Vauquelin, 75005 Paris, France}

\abstract{We study experimentally the dynamics of a dense intruder
  sinking under gravity inside a vibrated $2\mathcal{D}$ granular
  packing. The surrounding flow patterns are characterized and the
  falling trajectories are interpreted in terms of an effectivive
  friction coefficient related to the intruder mean descent velocity
  (flow rules). At higher confining pressures i.e. close to jamming, a
  transition to intermittent dynamics is evidenced and 
  displays anomalous "on-off" blockade statistics. A systematic
  analysis of the flow rules, obtained for different intruder sizes,
  either in the flowing regime or averaged over the flowing and
  blockade regimes, strongly suggest the existence of non-local
  properties for the vibrated packing rheology.}
\begin{document}
\maketitle 

%--------------------
\section{Introduction} 
During the last decades, several studies have shown amazing analogies
among variety of very different systems \cite{LN98} (colloids
\cite{Habdas2004}, foams \cite{Khan1988}, granular materials
\cite{GDR04,Majmudar2007}, supercooled liquids \cite{Cavagna2009}).
 Indeed, close to the jamming transition
these systems share common dynamical behaviors such as a glassy phase,
aging, memory or intermittency\cite{LN2001}. Based on
these analogies, several propositions emerged aiming to describe and
possibly understand, in a general and unified way, this phenomenology
and the associated rheology \cite{OHern2002,Bocquet2009}. In the case of dense granular packing, the rheology under
shear is described with good accuracy by local relations between a
dynamical friction coefficient and a local dimensionless number,
involving shear and confining pressure \cite{GDR04}: the inertial
number.  However, the boundary between fluid-like and solid-like
phases is still poorly understood. Experiments on avalanche flows
suggest that the rheology could indeed be non-local near arrest
\cite{Deboeuf2006, Aranson2006}. Conceptually, for the rheology of vibrated granular packing, the most
elaborated point of view stems from kinetics theory
\cite{Brey2001}. Whereas this simple vision is reliable to describe
dilute systems, it is severely challenged for dense systems \cite{Reis2007,Keys2007, Lechenault2008,Melhus2009}.
 It has been suggested in this case, that the notion of thermal
temperature could be replaced by the more general concept of
"effective temperature" describing the structural evolution between
blocked configurations \cite{MK02}; however,
practically, it is quite difficult to apply this concept in order to
obtain a clear constitutive picture. In experiments, a complication usually comes from the way energy is
input in the packing i.e. from the boundaries. It often leads to very
inhomogeneous bulk agitations, both in time and space. The
probes themselves are hard to decouple from the agitation energy input
\cite{Zik1992,Danna2003}, which may alterate the reliability of the
constitutive relations hence extracted\cite{Danna2003}. This was
somehow circonviened recently by Reis et al.\cite{Reis2007} and
Lechenault et al.\cite{Lechenault2008}, who studied $2\mathcal{D}$
horizontal packing driven homogeneously from the bottom; same for Keys
et al.\cite{Keys2007} using an hydrodynamic fluidization technique. In
these experiments, which approach jamming by increasing the packing compaction,
strongly heterogeneous and cooperative displacement modes  were
evidenced, very reminiscent to those present in undercooled liquids or
glassy systems\cite{Cavagna2009}.
\newline
\indent Recently, Caballero et al. \cite{Caballero09} studied a
sonofluidized $3\mathcal{D}$ granular packing and monitored the motion
of various intruders inside. Even at weak driving acceleration, the
resulting friction coefficients decrease significantly with
acceleration. Eventually, the Coulomb threshold vanishes and one
obtains a linear relation between force and velocity. These relations
are strongly dependent on the shape and the size of the intruders.
Other means of activation were
proposed using shear bands noise as actuators
\cite{Reddy2010,Nichol2010}. In these last cases, empirical relations
between force and velocity were extracted and all these works
concluded on the non-local character of the constitutive relations.
However, in all these cases, the intruder was buried partially or completely in
the bulk, thus hindering a direct visualization of granular motion in
the intruder surrounding.

In the present work, we seek to extract the rheological properties of
a vibrated vertical granular packing by monitoring the vertical motion
of cylindrical intruders. This approach is very analogous to bead
rheometers techniques currently used in fluid mechanics. Note also
some recent work along those lines on intruder penetration in
$2\mathcal{D}$ or $3\mathcal{D}$ packing but in the absence of
vibration \cite{Albert1999,Geng2004}. This kind of experiments were
also performed in other systems like colloids \cite{Habdas2004},
glasses \cite{Hastings2003} or foams \cite{Dollet2005_2007}.
 
\indent The paper is organized as follows. First, we present the
experimental setup, a $2\mathcal{D}$ experiment realization of a
vibrated granular packing oriented in the direction of the gravity.
Then, the descent dynamics of heavy intruders are monitored in
association with the resulting counterflow of the surrounding
particles. We show that the analysis of the intruder trajectories
may probe a rheological behavior via a relation between an
effective friction coefficient and the mean intruder descent velocity. We
analyze the intermittent statistics evidenced at higher depths as an
alternation of flow and blockade. We finally question the existence
of a local relationship accounting for the rheology of the dense
granular medium under vibration.
%
%-----------------------------------------
\section{Set-up and experimental protocol}
\begin{figure}[h!]
\centering
\includegraphics[width=0.37 \textwidth]{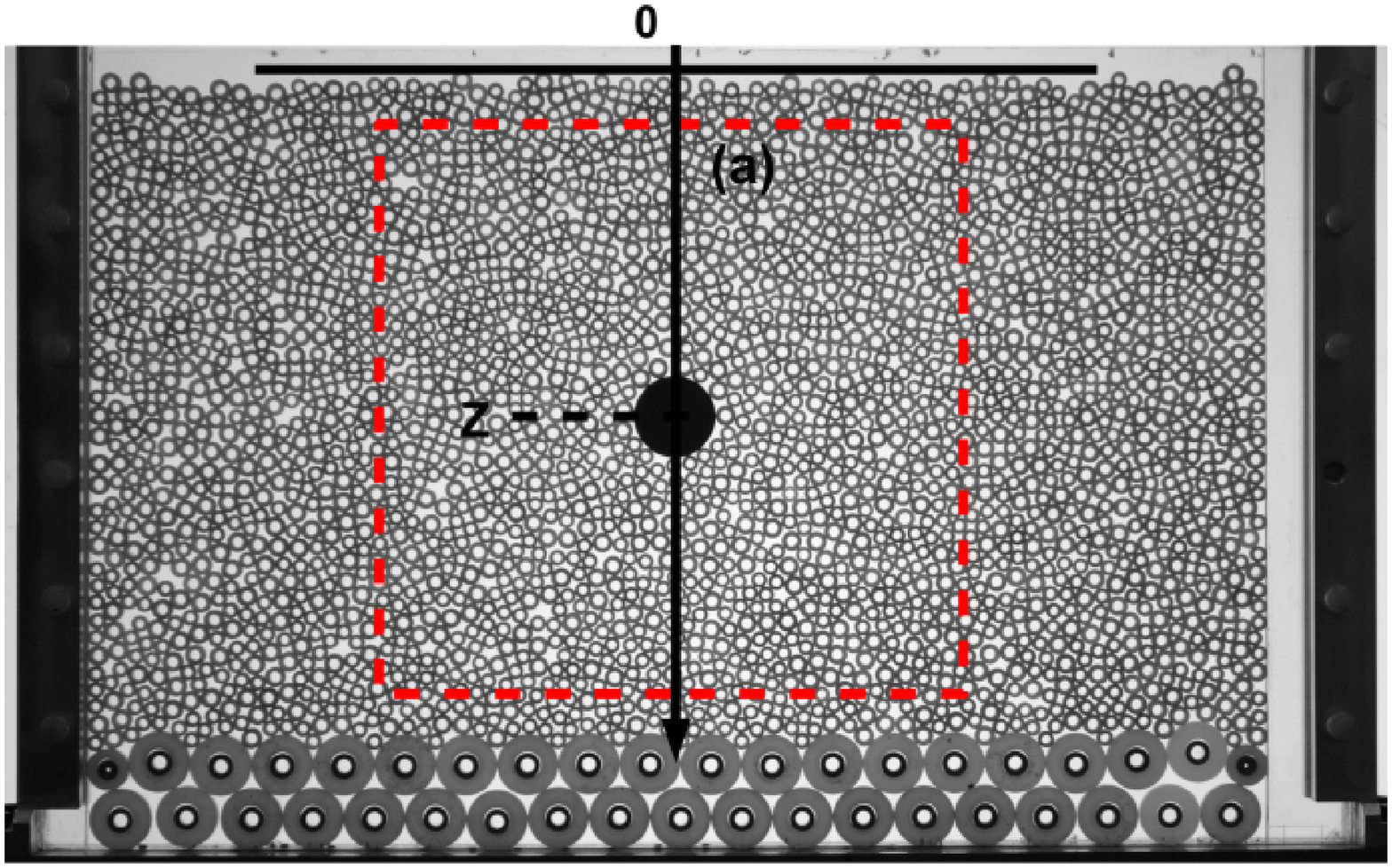}%0.25
\\
\includegraphics[width=0.4 \textwidth]{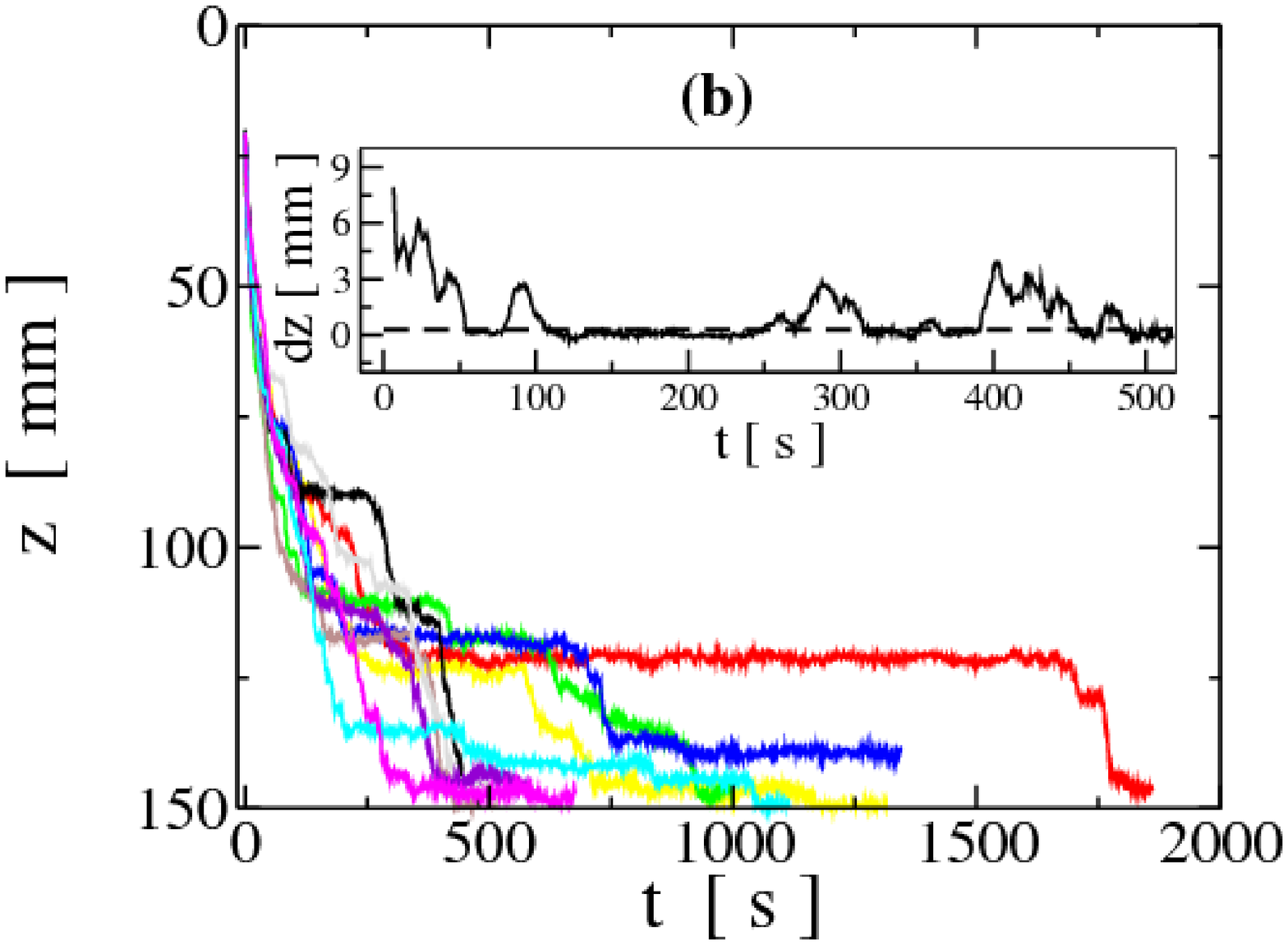}
\caption{(a) View of the cell containing the bi-disperse packing and a
  cylindrical intruder of radius $R = 10$ mm with $\rho \cong 0.62~g.cm^{-3}$. The
  dashed rectangle is the visualization field. (b) Vertical positions
  of the intruder as a function of time for $10$ experiments with
  identical conditions of vibrations (radius $R = 10$ mm, density
  contrast $\frac{\Delta\rho}{\rho} = 16.15$). Inset : vertical
  increment of displacement for a time lag 1s for one of the
  intruder trajectories presented in the main graph (same color).}
\label{Figure1}
\end{figure}

Figure \ref{Figure1}(a) displays a snapshot of the experimental
set-up. The cell is a vertical rectangular container of lateral size
$L = 275$ mm and thickness $3.5 $ mm slightly larger than the
thickness of the disks forming the model granular packing. We use a
bi-disperse mixture of around $1100$ big (diameter $d_b = 5$ mm) and
$1600$ small (diameter $d_s = 4$ mm) nylon disks of thickness $3$
mm. The intruder is a metallic cylinder of adjustable radius $R$ and
same thickness as the grains. The average intruder density can also
be adjusted by drilling a hole at its center. The density contrast
with the surrounding vibrated granular medium of density $\rho$ is
defined as $\frac{\Delta \rho}{\rho}$ and is always chosen
positive. Therefore, the intruder will sink due to buoyancy if the
surrounding medium is fluidized enough.
\newline
\indent The way vibration energy is input in the system, differs from
what has been done in previous experiments. Here, we seek to avoid
synchronization between the global motion of the packing and the
sinusoidal motion of the driving plate. It is known that for
accelerations larger and around gravity $g$, the packing trajectory
and the energy input due to repetitive collisions with the bottom
plate, are strongly synchronized, unless a very high acceleration
level is reached (up to $30g$!\cite{Eshuis05}). Note also that for
driving accelerations larger than $g$, surface instabilities such as
Faraday waves may also occur which render the packing spatially
inhomogeneous \cite{Clement96}. Here, the idea is to break as much
as possible this spatio-temporal synchronisation responsible for the
alternation of phases at drastically different agitation levels.
To this purpose, we place a so-called "vibrating chain" at the bottom
of the cell (see Figure \ref{Figure1}(a) bottom) which consists of two
arrays of large and dense disks activated from below by $24$
electromagnetic pistons oscillating vertically with an amplitude $A_p$
inside the granular packing and thus, transferring vibration energy to
this "chain". A detailed study of bulk agitation and packing density
hence obtained at steady-state, is discussed elsewhere
\cite{thesis_harich}.
\newline
\indent In this report, the packing is vibrated under fixed driving
conditions : piston amplitude $A_P = 6$ mm and driving frequency $f_D
= 20$ Hz with a phase shift of $\pi$ between neighboring pistons. We
measured for this system a driving RMS acceleration of $2g$ in the
vertical direction. In a typical experiment, the vibration is
activated $15$ minutes before we place the intruder at the piling
surface. Its descent is monitored by a CCD camera taking pictures
every second. For each intruder we perform $10$ experiments in similar
experimental conditions.

%---------------------------------  
\section{Intruder descent dynamics} 
\begin{figure*}[htp]%[h!]
\centering
\includegraphics[width=0.26 \textwidth]{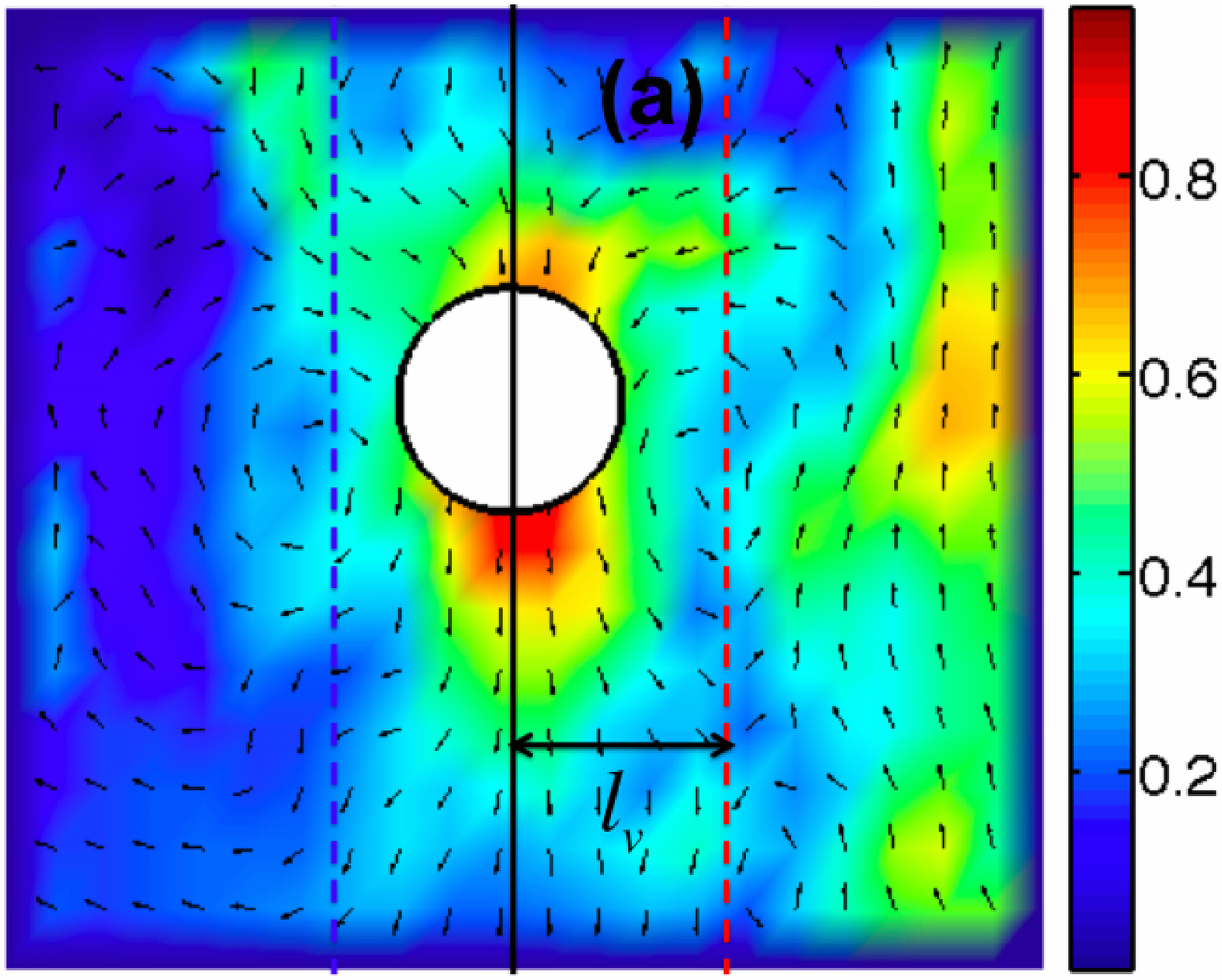}
\includegraphics[width=0.3 \textwidth]{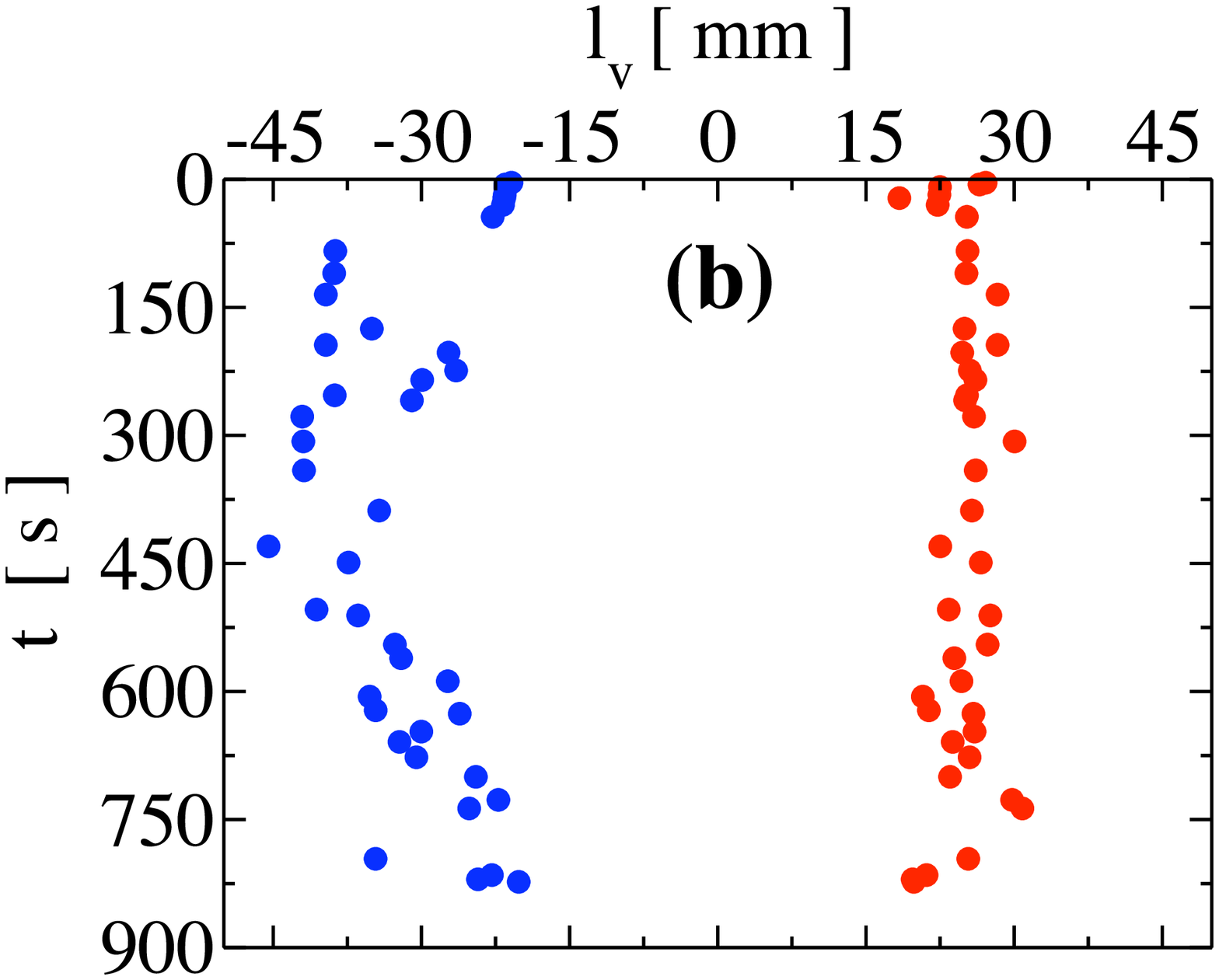}
\includegraphics[width=0.32 \textwidth]{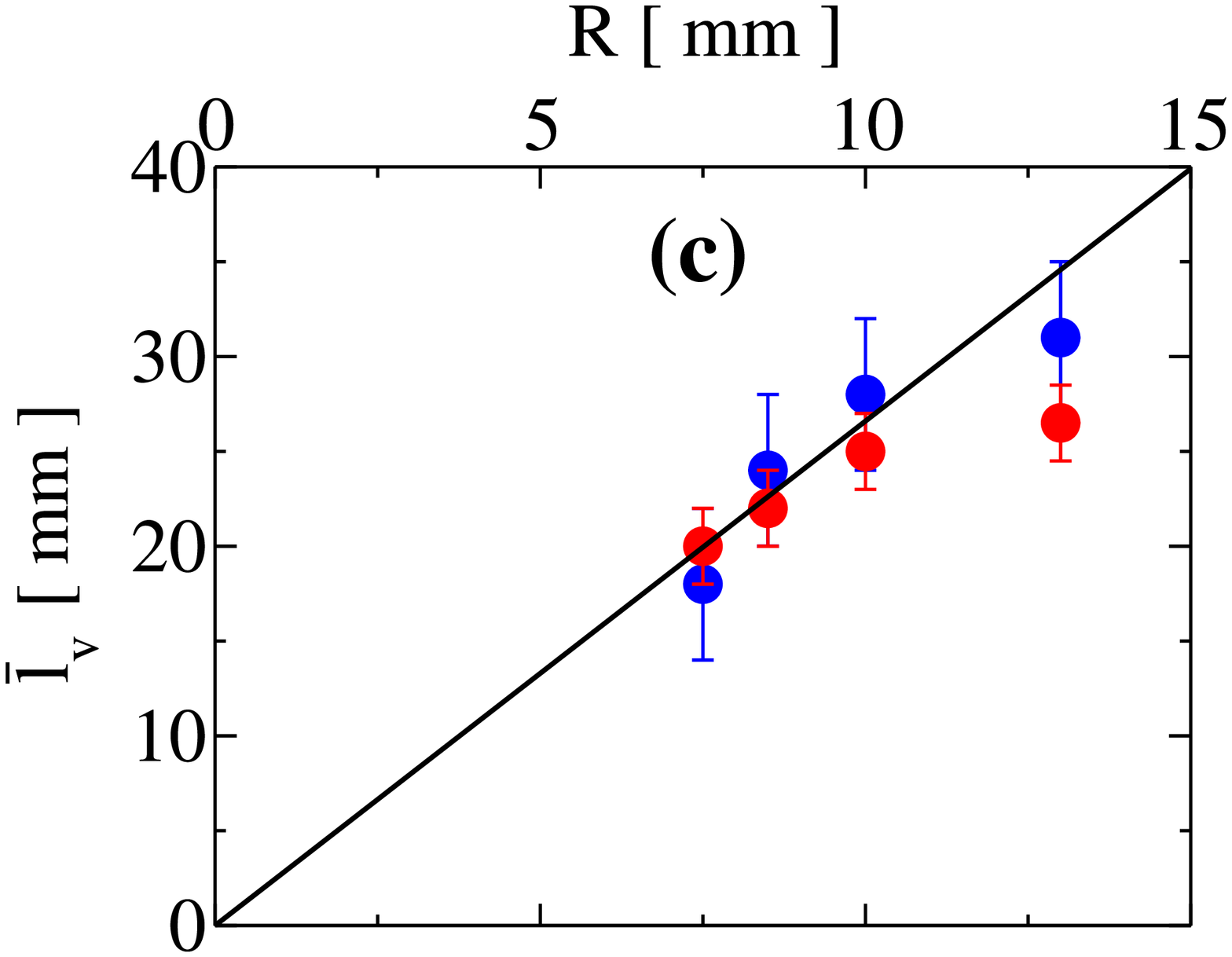}
\caption{(a) Average displacement field for a time lag corresponding
  to an intruder displacement over its radius $R = 10$ mm and for a
  coarse-graining size $\Delta W=d_s$. The color code represents the
  amplitude of the coarse-grained displacement normalized by the
  intruder radius and the vectors indicate the direction of grain
  motion. Note the horizontal distance $l_v$ between the
  intruder center and a vortex center. (b) Horizontal positions of
  left and right vortex centers as a function of time (
  density contrast $\frac{\Delta\rho}{\rho}$=16.15),
  ((\textcolor{red}{$\bullet$}) right vortex and
  (\textcolor{blue}{$\bullet$}) left vortex). (c) Mean horizontal
  distances $\bar{l}_v$ between the intruder and the vortex centers as
  a function of the radius $R$ (symbol colors as in
  (b)).}
\label{Figure3}
\end{figure*}

Figure \ref{Figure1}(b) shows the vertical positions $z(t)$ of an
intruder of radius $R = 10$ mm as a function of time. The first
noticeable feature is that the intruder motion is quite irregular with
alternate phases of flow and blockade. The deeper the intruder is in
the packing, the more likely blockade seems to occur (see inset of
figure \ref{Figure1}(b)). This intermittent dynamics yields a large
distribution of arrival times at the bottom of the cell as observed
from the different vertical trajectories taken in identical
conditions.
\newline
\indent Figure \ref{Figure3}(a) displays the coarse-grained
displacement field over a time lag $\tau$ defined using the mean
intruder sinking time over a distance $R$. A feature of these
displacement fields is the existence of two vortices accompanying the
intruder's descent and corresponding to recirculations of grains
around the intruder. Note that during a descent, one of the vortices
seems to be dominant in amplitude but it can be located on either side
of the intruder. In figure \ref{Figure3}(b), the distances $l_v$
joining the intruder's center to both vortex centers are represented
as a function of time.
From Figure \ref{Figure3}(b), we see that during the whole fall both
vortices seemed to stay at a fixed distance $\bar{l}_v(R)$ from the
intruder. We thus represented this mean
distance $\bar{l}_v(R)$ for experiments done with different intruder
sizes and observed that these distances scale almost linearly with the
radius $R$ ( Fig. \ref{Figure3}(c)).
%
%---------------------------------------------------------------
\section{Effective friction}
One objective of this study is to shed light on the rheological
properties of granular matter under vibration. In either static,
quasi-static or dense flow situations, the central constitutive
parameter describing the granular rheology is the ratio $\mu$ between
shear stress and confining pressure (see \cite{GDR04} and
refs. inside), i.e., a Coulomb friction coefficient. Scaling analysis
shows that for an intruder of radius $R$, the driving force per unit
of thickness $F_{G}$ is set by gravity, thus $F_{G}\varpropto \Delta \rho
gR^{2}$. This force must be balanced at steady state by the resulting
friction force exerted on the intruder boundary $F_{S}\varpropto \mu
\rho gzR$. Thus, a relation can be obtained for the friction
coefficient $\mu$ which scales like $\mu \varpropto \Delta \rho R/\rho
z$, where $\frac{\Delta \rho}{\rho}$ is the density contrast, $R$ the
intruder radius and $z$ its vertical depth. This argument can be made
more quantitative in the framework of a simple reference model
involving a constant dynamical friction $\mu$ so that effective
friction and dynamical friction would match ($\mu_e=\mu$). The force
balance can be computed exactly by considering a local confining
pressure imposed by gravity $P=\rho gz$ and by integrating the
friction force around the intruder. This is the base for our
definition of an effective friction coefficient $\mu_e$:
\begin{equation}
\mu_{e} = \frac{\pi}{4} \frac{\Delta \rho}{\rho}\frac{R}{z}
\end{equation}
\newline
\indent This relation motivates a parametric study of the intruder
motion by varying its size and its apparent density. For each
realization $i$, we measure the time $\tau_{i}(z)$ spent by the
intruder to span the interval $[z-\Delta z/2,z+\Delta z/2]$. Then, the
mean descent velocity is defined as $\bar{V}(z)=<\Delta z/\tau
_{i}(z)>_{i}$, the average being taken over $10$ realizations.

%--------------------------------------
\section{Dimensionless number} 
\begin{figure}[h!]
\centering
\includegraphics[width=0.4 \textwidth]{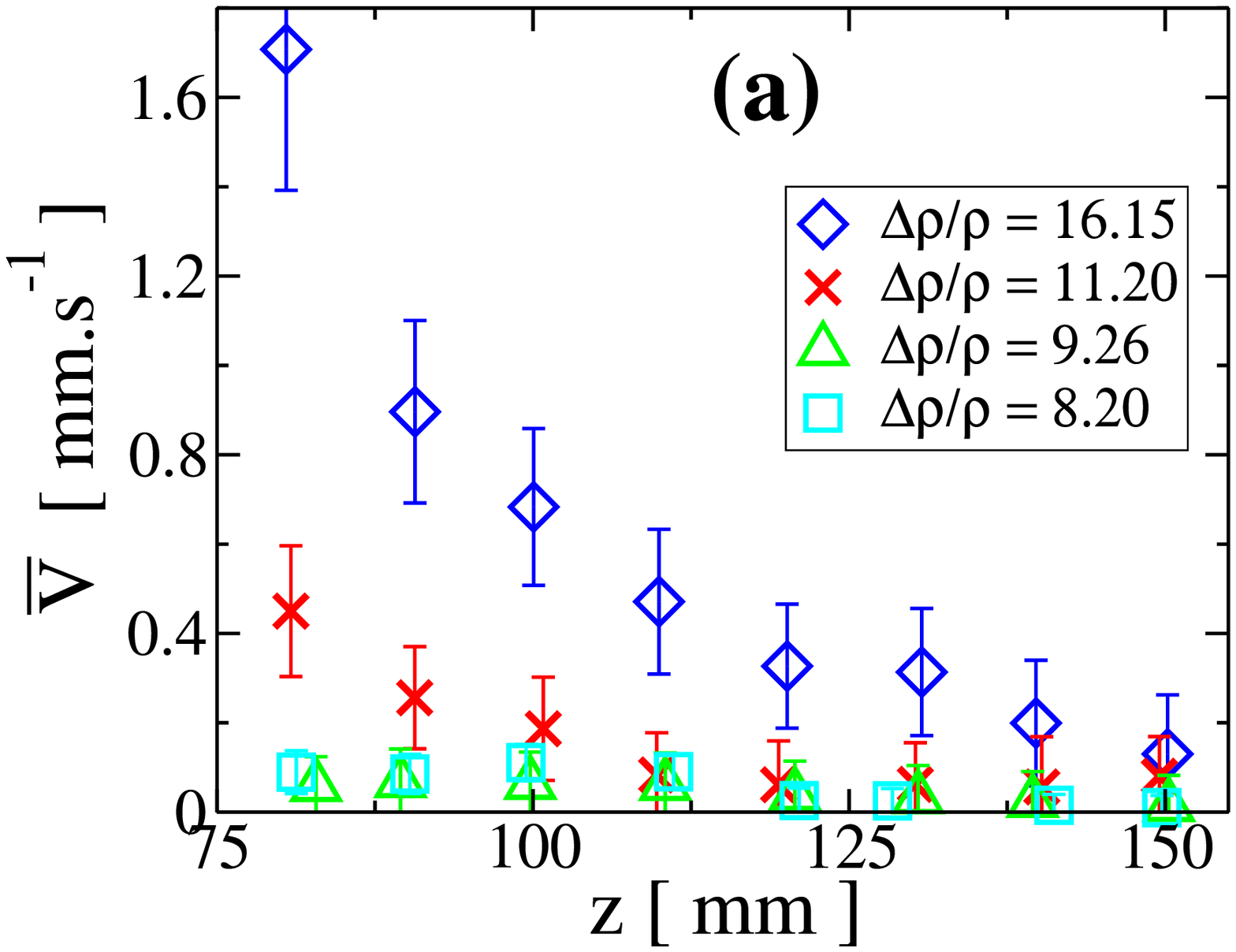}
\\
\includegraphics[width=0.4 \textwidth]{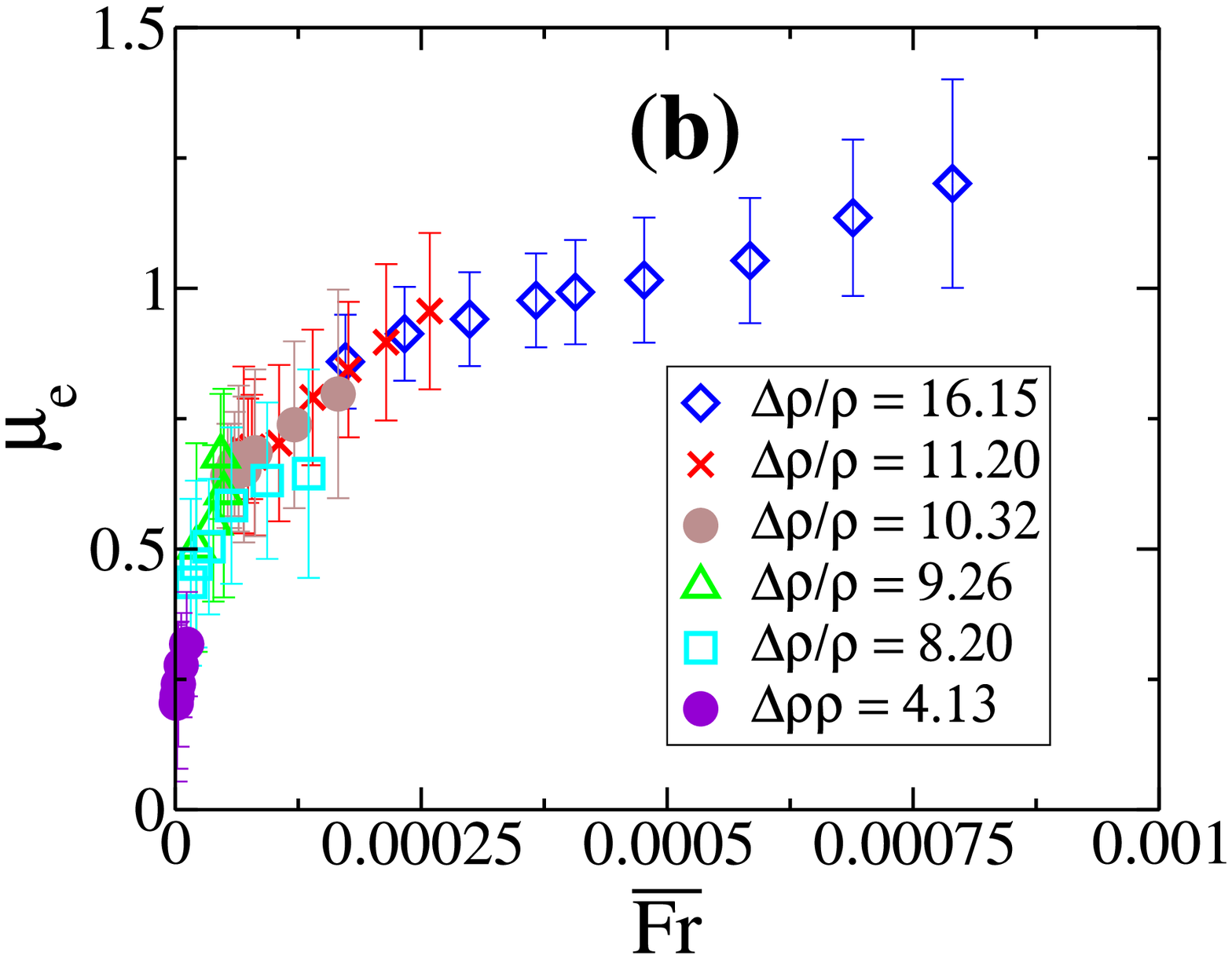}
\caption{(a) Mean velocity $\bar{V}(z)$ as a function of depth $z\pm
  \Delta z/2$ with $\Delta z = 20$ mm and for $R = 10$ mm and
  different density contrasts $\frac{\Delta\rho}{\rho}$. (b) Effective
  friction $\mu_e(z)$ as a function of Froude number
  $\bar{Fr}(z)={\bar{V}\over \sqrt{P/\rho}}$ for different density
  contrasts.}
\label{Figure4}
\end{figure}

In figure \ref{Figure4}(a), we display for an intruder of radius $R =
10$ mm, the mean descent velocity $\bar{V}(z)$ obtained for $\Delta z
= 20$ mm and for different density contrasts.  As expected the mean
velocity increases with density contrast. We also note the very
regular decrease of mean velocity with depth $z$.  In order to rescale
the different curves, we define for each depth $(z)$, a local mean
Froude number as $\bar{Fr}=\bar{V}/\sqrt{P/\rho}=\bar{V}/\sqrt{gz}$.
In Figure \ref{Figure4}(b), we observe a rescaling of all the data on
a common flow rule, which justifies the existence of an effective
friction coefficient describing the rheology of the granular
intrusion. Interestingly, this flow rule is of the hardening type,
i.e., the effective friction increases with the local drag
velocity, a behavior also observed for sheared dense granular
media\cite{GDR04} without \cite{Geng2004} or with vibration \cite{Caballero09, Reddy2010,Nichol2010}.
%
%------------------------------------ 
\section{Blockade and flow statistics} 
\begin{figure}[h!]
\centering
\includegraphics[width=0.4 \textwidth]{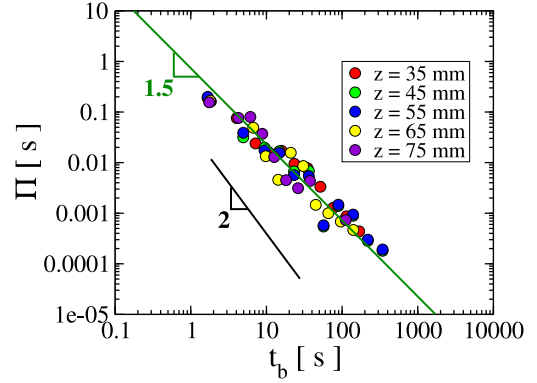}
\caption{Probability density function of blockade times $\Pi(t_b)$
  for $R = 10$ mm and $\frac{\Delta\rho}{\rho}=16.15$, at different
  depths.}
\label{Figure5}
\end{figure}

We previously noticed (see Figure \ref{Figure1}(b)) that a prominent
feature of the intruder descent dynamics is the intermittent sequel of
blockades and flows. This clearly has a strong impact on the effective
mean velocity as measured previously. In the following, we will seek
to separate and characterize both regimes independently. The method we
design to extract the blocked phases is based on the use of a
displacement threshold $\delta_{Z}$ at a time scale $\delta t = 1$ s
below which we consider the intruder is blocked (see the dashed line
in inset of figure \ref{Figure1}(b)). This choice is somehow
arbitrary; however, we rationalized it by defining, for each descent
curve, a value such that the jump distribution in the blocked phase is
close to a symmetric Gaussian curve with a zero average.
 For each trajectory, we identify the depths at which the
intruder gets blocked and starts flowing again as well as the time
spent in both phases respectively. Figure \ref{Figure5} displays the
distribution of times spent in the blocked phase $t_b$ at different
depth values ($z\pm \Delta z/2$). The striking feature is the outcome
of a power-law distribution $t^{-\alpha}$, with an exponent smaller
than $2$ ($\alpha \cong 1.5$). Such a distribution is the sign of an
anomalous statistics as no typical time scale can be defined. To
characterize the sojourn in the blocked phase, we thus consider the
fraction of time $\Phi_{b}$ spent in the blocked phase at different
depths.
\begin{figure}[h!]
\centering
\includegraphics[width=0.4 \textwidth]{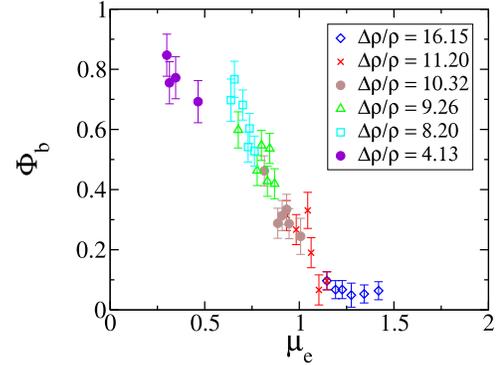}
\caption{Fraction of time $\Phi_b(z)$ spent in the blocked phase as a
  function of the effective friction $\mu_e(z)$ for $R = 10$ mm and
  different density contrasts $\frac{\Delta\rho}{\rho}$.}
\label{Figure6}
\end{figure}
\newline
\indent In figure \ref{Figure6}, $\Phi_{b}$ is plotted as a function
of the effective friction $\mu_{e}(z)$. At low $\mu_e$, $\Phi_b$ is
large and almost constant. At $\mu_{e}\cong 0.7$ a sharp decrease of
the fraction of blocked time is observed. This transition from
blockade to flow suggests a generalization of the static Coulomb
threshold in the presence of vibration. Note that this effect of threshold distribution was also evidenced in numerical simulations of sheared granular packings under random noise \cite{Melhus2009}.\\
Now we want to characterize the flowing phases. We define a flowing
time variable $t_F$ by removing \emph{sequentially} the blockade periods of
the trajectories. Figure \ref{Figure5b}(a) displays the flowing
trajectories i.e. the intruder positions as a function of $t_F$. A
clear collapse of all the trajectories is observed and an average is
performed over $10$ experiments done in identical conditions in order
to define a mean flowing trajectory $z(t_F)$. A flowing velocity is
then computed at each depth $\bar{V}_F={dz \over dt_F}$. This
procedure allows to define a new Froude number $\bar{Fr}_F={\bar{V}_F
  \over \sqrt{P/\rho }}$, (with the subscript F for flow) that we call the dynamical Froude number as
it applies to the flowing phase only.
\newline
\indent In figure \ref{Figure5b}(b), we display the relation $\mu
_{e}(\bar{Fr}_F)$ for an intruder of radius $R = 10$ mm and for various
density contrasts. Interestingly, we also observe a data collapse onto a
unique curve.
\begin{figure}[h!]
\centering
\includegraphics[width=0.4 \textwidth]{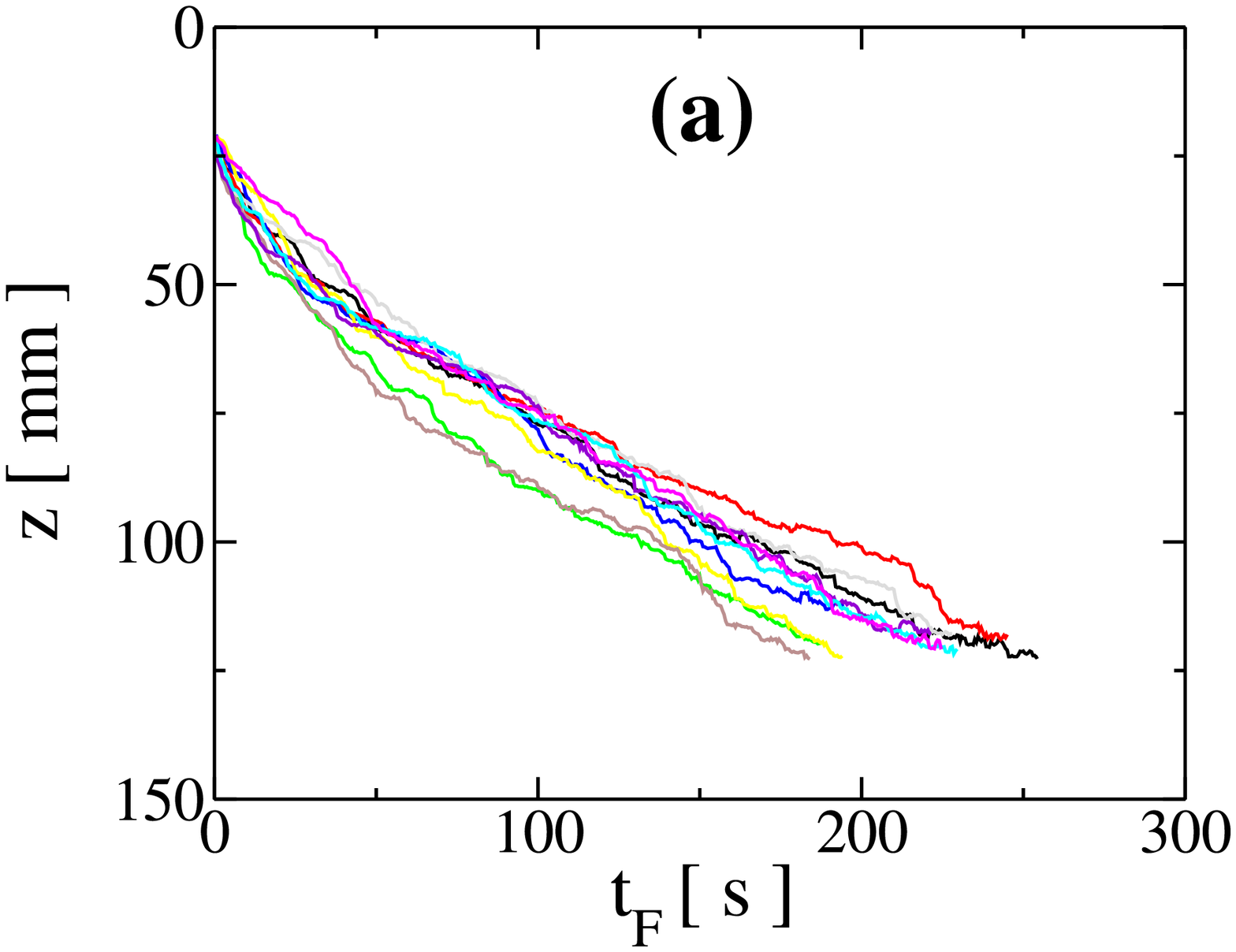}
\\
\includegraphics[width=0.4 \textwidth]{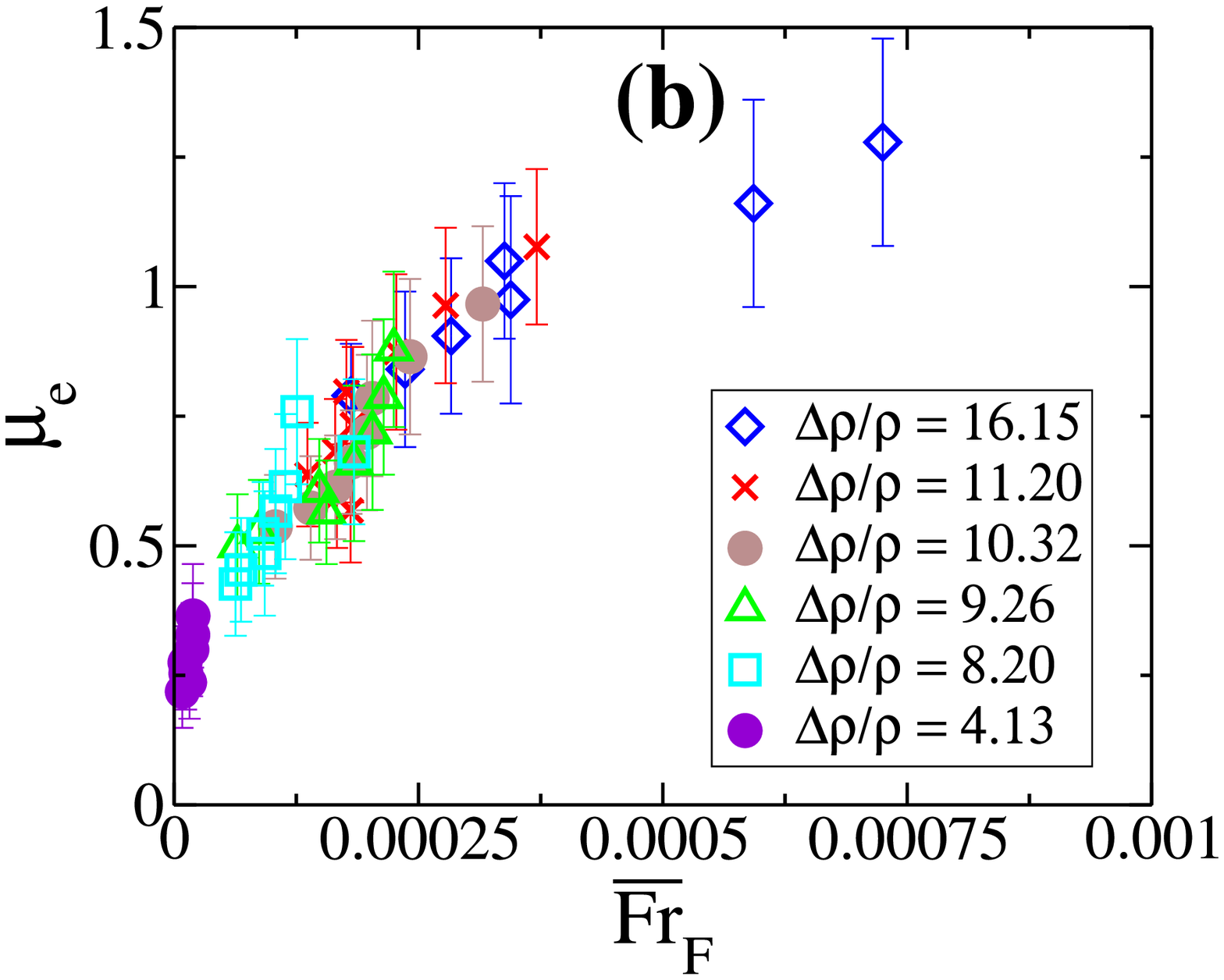}
\caption{(a) Vertical positions $z$ of the intruder with $R = 10$ mm corresponding to
  fig.\ref{Figure1}(b), as a function of the flowing time variable
  $t_F$. (b) Dynamical flow rule : effective friction $\mu_e(z)$ as a
  function of the dynamical Froude number
  $\bar{Fr}_F(z)={\bar{V}_F(z)\over \sqrt{gz}}$ for various density
  contrasts $\frac{\Delta\rho}{\rho}$.}
\label{Figure5b}
\end{figure}
%

%----------------------------------
\section{Towards a non-local rheology ?} 
\begin{figure}[h!]
\centering
\includegraphics[width=0.4 \textwidth]{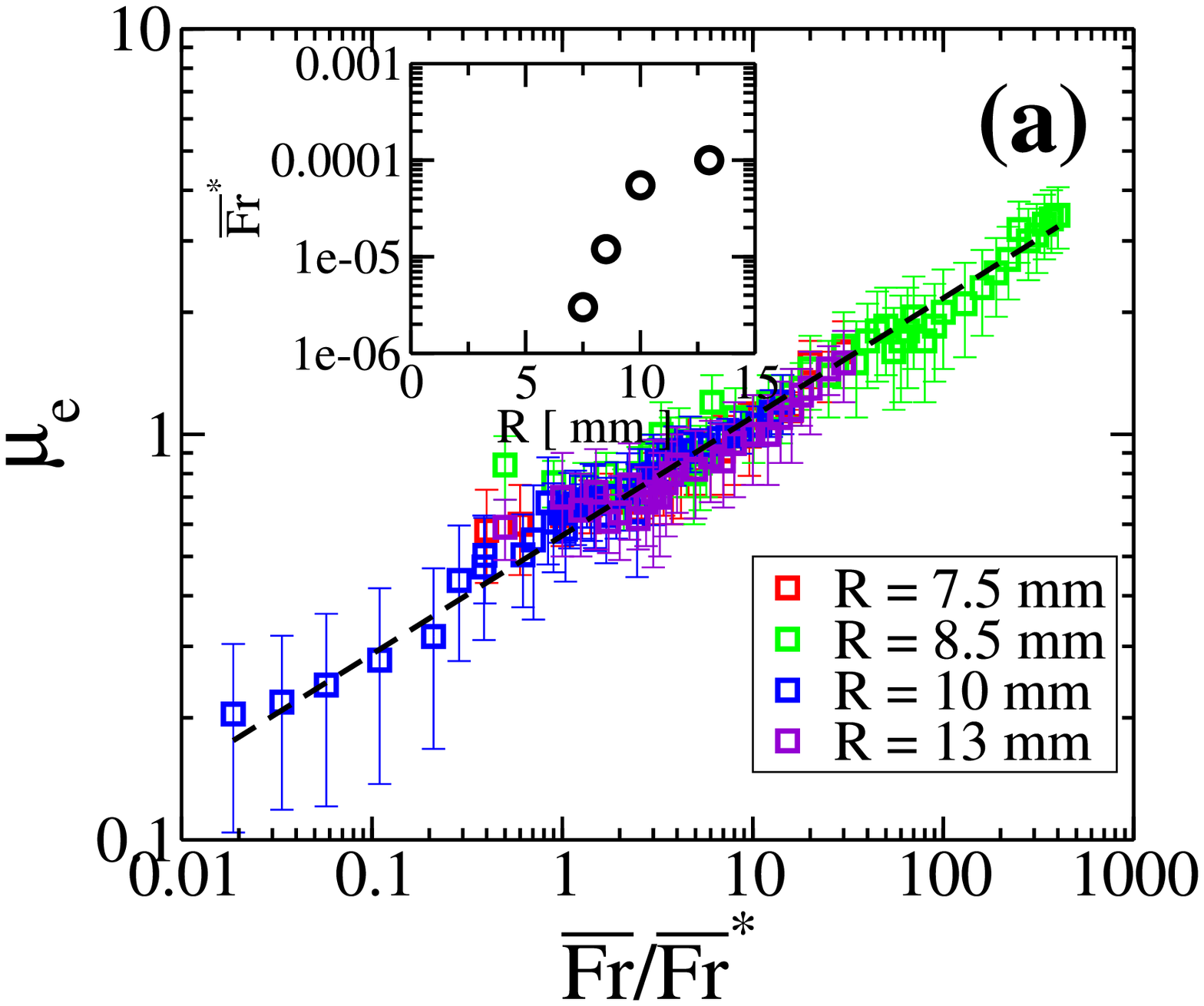}
\\
\includegraphics[width=0.4 \textwidth]{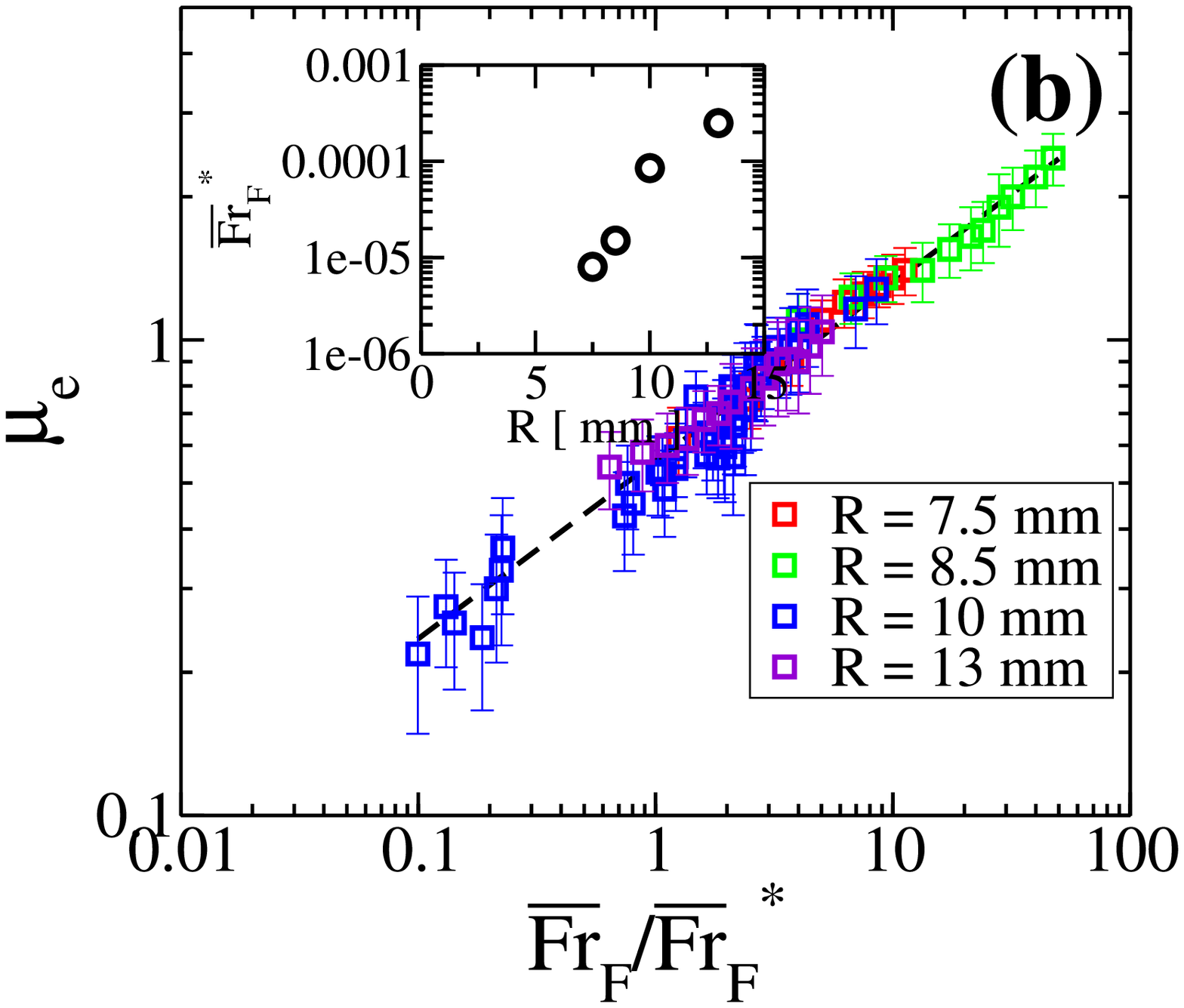}
\caption{(a) Mean flow rules for different intruder sizes represented
  as a function of a rescaled Froude number $\bar{Fr}\over
  \bar{Fr}^{*}$ for different intruder radius $R$ (fit $A_0.x^{A_1}$
  with $A_0=0.56$ and $A_1=0.29$). Inset : rescaling Froude number
  $\bar{Fr}^{*}$ as a function of the intruder radius. (b) Dynamical
  flow rule obtained in the flowing phase as a function of a rescaled
  Froude number ${\bar{Fr}_f\over \bar{Fr}_F^{*}}$ (fit $A_0.x^{A_1}$
  with $A_0=0.56$ and $A_1=0.37$). Inset : rescaling Froude number
  $\bar{Fr}_F^{*}$ as a function of the intruder radius. Note the log
  scales for the vertical axis.}
\label{Figure7}
\end{figure}

To extract a local rheology that could reflect constitutive properties
of granular matter under vibration, we must at the end, provide a
relation between a constitutive coefficient (friction) and local
macroscopic fields such as pressure or shear rate and this,
independently of the intruder's presence. To address explicitly this
question, we computed systematically the mean and dynamical flow rules
corresponding to intruders whose diameter varies between $3.75$ and $6.5$
small granular sizes $d_s$. We obtained in each case - like what was
presented previously for $R = 10 $mm - a rescaling of the data with
different density contrasts. By visualization of the intruder flows
(see Figures \ref{Figure3}(a) and (b)), we have seen that shear is
exerted on the intruder side at a distance of the order of $l_v(R)$
yielding a typical shear rate $\dot{\gamma}\cong{\bar{V}\over
  l_v(R)}$. Following the analysis provided for rapid dense flows, we
may try to transform the flow rules into a local rheology law using an
inertial parameter $I={\dot{\gamma }d_s\over
  \sqrt{P/\rho}}={d_s\bar{V}(z)\over \l_v\sqrt{gz}}={d_s\over
  \l_v}\bar{Fr}$. If a local rheology applies, a relation $\mu_e (I)$
should show up independently of the intruder size. We see in Figures
\ref{Figure7}(a) and (b) that the flow rules for different intruder's
size can be rescaled by dividing the Froude numbers by a value
$\bar{Fr}^{*}(R)$ for the mean flow rule or $\bar{Fr}_F^{*}(R)$ for the dynamical flow rule. 
These rescaling values are presented as a
function of the intruder radius $R$ in insets of Figures
\ref{Figure7}(a) and (b). However, we observed that for both flow
rules the corresponding lengths $\bar{Fr}^{*}(R)d_s$ and
$\bar{Fr}_F^{*}(R)d_s$, grow much faster with $R$ than
$l_v(R)$. Therefore these
rescalings could not provide a value such that the Froude ratios could
be identified with an inertial parameter independent of $R$.
Thus, it becomes clear that contrarily to the situation of rapid
dense flows, the flow rule cannot be transformed into a local
rheology using an inertial parameter. Whether this is the sign of
true non-locality in the constitutive relations associated with a growing
length scale near jamming or the unravelled influence of a supplementary field such as "fluidity" or " granular temperature"
is left for future discussions. 
Note finally that under constant velocity and confining pressure, the effective friction
increases when the intruder size decreases. This is the so-called
"geometrical hardening" effect already
noticed by Caballero et al.\cite{Caballero09}.
%
%------------------   
\section{Conclusion}
In this experimental study, we measured the dynamics of dense
intruders sinking in a vibrated granular packing. The flow fields
around the intruder was measured and characterized by two side
vortices accompanying the intruder's descent. The mean position of
these vortices scales with the intruder size. The sinking dynamics is 
interpreted as as an effective friction
coefficient increasing with the mean falling velocity. For a given
intruder size, when the mean falling velocity $\bar{V}(z)$ is rescaled
as a Froude number $Fr(z)= {\bar{V}(z) \over \sqrt{P(z)/\rho}}$, where
$P(z)$ is the local confining pressure, a rescaling of all the flow
curves is evidenced, for different intruder densities. This defines
the so called "mean flow rule" characterizing the intruder rheology.
An important feature is that at large confining pressures, the
intruder dynamics is intermittent and displays alternating phases of
blockade and flow. The time spent in the blocked phase follows a
statistics with a decaying power-law displaying an anomalous exponent
around $1.5$. This is quite reminiscent of the "on-off" intermittency
process \cite{AumaitreJSP2006} where the blocked phases show up with a
similar residence time distribution, also evidenced for the flow out of a vibrated
hopper\cite{Janda2009}. We interpret the transition between blockade
and flow as an effective friction threshold appearing as a "noisy"
generalization of the Coulomb threshold.
When isolated, the dynamics of the flowing phases, displays a similar
data collapse defining a "dynamic flow rule", however with a different
functional form as for the "mean flow rule".
Finally, for different intruder sizes, we could collapse both flow
rules using a rescaled value of the Froude number increasing exponentially with
the intruder size. We discuss the fact that this very strong
dependance might be an indication that the effective rheology of the
agitated packing is non-local.\\
%*******************************
\textit{Acknowledgement}: We thank Bruno Andreotti for many scientific
discussions and the CNES-2010 and ANR-2010-JamVibe programs for
financial support.


\begin{thebibliography}{}
\bibitem{LN98} A. Liu, S.R. Nagel, Nature \textbf{396}, 21 (1998).

\bibitem{Habdas2004} P. Habdas, D. Schaar, A. Levitt and E. Weeks,
  Europhys. Lett. \textbf{67}, 477 (2004).
  
\bibitem{Khan1988} S. Khan, C. Schnepper and R. Amstrong,
  J. Rheol. \textbf{32}, 69 (1988).
  
\bibitem{GDR04}GDR Midi, Eur. Phys. J. E, \textbf{14}, 341 (2004);
  F. da Cruz, S. Emam, M. Prochnow, J.-N. Roux, and F.  Chevoir,
  Phys. Rev. E \textbf{72}, 021309 (2005).

\bibitem{Majmudar2007} T.S. Majmudar, M. Sperl, S. Luding and
  R.P. Behringer, Phys. Rev. Lett. \textbf{98}, 058001 (2007).
  
\bibitem{LN2001} A. Liu and S.R. Nagel (Editors), Jamming and
  rheology: Constrained Dynamics on Microscopic and Macroscopic Scales
  (Taylor \& Francis, New York) 2001;  L. Berthier and G. Biroli, Encyclopedia of
  Complexity and Systems Science (Springer, New York, 2008).

\bibitem{OHern2002} C. O'Hern, S. Langer, A. Liu and S.R. Nagel,
  Phys. Rev. Lett. \textbf{88}, 036001 (2002).
  
\bibitem{Bocquet2009} L. Bocquet, Phys. Rev. Lett. \textbf{103},
  036001 (2009).

\bibitem{Deboeuf2006}S. Deboeuf, E. Lajeunesse, O. Dauchot, and
  B. Andreotti, Phys. Rev. Lett. \textbf{97}, 158303 (2006).

\bibitem{Aranson2006} I. S. Aranson, L. S. Tsimring, F. Malloggi, and
  E. Cl\'ement, Phys. Rev. E,\textbf{78}, 031303 (2008).

\bibitem{Brey2001} J. J. Brey, M. J. Ruiz-Montero, and F. Moreno,
  Phys. Rev. E,\textbf{63} 061305, (2001).

\bibitem{MK02} H. A. Makse and J. Kurchan, Nature \textbf{415}, 614
  (2002); F. Q. Potiguar and H. A. Makse,
  Eur. Phys. J. E, \textbf{19}, 171 (2006).

\bibitem{Zik1992} O. Zik, J. Stavans, and Y. Rabin, Europhys.
 Lett. \textbf{17}, 315 (1992).

\bibitem{Danna2003} G. D'Anna et al., Nature \textbf{424}, 909 (2003).

\bibitem{Reis2007} P. M. Reis, R. A. Ingale, and
  M. D. Shattuck. Phys. Rev. Lett., \textbf{98} 188301, (2007).

\bibitem{Lechenault2008} F. Lechenault, O. Dauchot, G. Biroli, and
  J. P. Bouchaud, Europhys. Lett.,\textbf{ 83} 46003, (2008).

\bibitem{Keys2007}A. S. Keys, A. R. Abate, S. C. Glotzer and
  D. J. Durian, Nature Physics, \textbf{3} 260 (2007).
  
\bibitem{Cavagna2009} A. Cavagna, Physics Reports, \textbf{476}, 51 (2009).
 
\bibitem{Caballero09} G. A. Caballero and E. Cl\'ement,
  Eur. Phys. J. E, \textbf{30} 4 (2009).

\bibitem{Reddy2010} K. A. Reddy, Y. Forterre, and O. Pouliquen,
  Phys. Rev. Lett. \textbf{106}, 108301 (2011).

\bibitem{Nichol2010}K. Nichol et al.,Phys. Rev. Lett. \textbf{104}, 078302 (2010).

\bibitem{CandelierPRL2009} R. Candelier and O. Dauchot,
  Phys. Rev. Lett. \textbf{103}, 128001 (2009).
  
\bibitem{Seguin2011}A. Seguin, Y. Bertho, P. Gondret, and J. Crassous,
  \textit{Dense granular flow around a penetrating object: Experiments and
  hydrodynamic model}, submitted (2011).

\bibitem{Hastings2003} M.B. Hastings, C.J. Olson Reichhardt and
  C. Reichhardt, Phys. Rev. Lett. \textbf{90}, 098302 (2003)

\bibitem{Albert1999}R. Albert et al., Phys. Rev. Lett. \textbf{82}, 205 (1999);
  I. Albert et al., Phys. Rev. Lett. \textbf{84}, 5122 (2000).

\bibitem{Geng2004}J. Geng, R.P. Behringer, Phys. Rev. Lett. \textbf{93}, 238002
  (2004); Phys. Rev. E \textbf{71}, 011302 (2005).

\bibitem{Dollet2005_2007} B. Dollet et al., Phys. Rev. E \textbf{71},
  013403 (2005).
  
\bibitem{Eshuis05}P. Eshuis, K. van der Weele, D. van der Meer and
  D. Lohse, Phys. Rev. Lett. \textbf{95}, 258001 (2005).
  
\bibitem{Melhus2009} M. F. Melhus, I. S. Aranson, D. Volfson, L. S. Tsimring, Phys. Rev. E \textbf{80},
  041305 (2009).
  
\bibitem{Clement96} E. Cl\'ement, L. Vanel, J. Rajchenbach, J. Duran,
  Phys. Rev. E \textbf{53} 2972 (1996).
  
\bibitem{thesis_harich} R. Harich, \textit{Doctoral thesis},
  University Pierre et Marie Curie (2010).
  
\bibitem{AumaitreJSP2006}S. Aumaitre, K. Mallick and F. Petrelis,
  J. Stat. Phys., \textbf{123}, 909, (2006).
  
\bibitem{Janda2009} A. Janda \etal, Europhys Lett, \textbf{87}, 24002 (2009).
%%  
\end{thebibliography}
\end{document}